# Dataspace architecture and manage its components class projection

## N. Shakhovska, Y. Bolubash

Information Systems and Networks Department,
Lviv Polytechnic National University
S.Bandera str, 28, Lviv, Ukraine,
natalya233@gmail.com

Abstract. Big Data technology is described. Big data is a popular term used to describe the exponential growth and availability of data, both structured and unstructured.
There is constructed dataspace architecture. Dataspace has focused solely – and passionately – on providing unparalleled expertise in business intelligence and data warehousing strategy and implementation. Dataspaces are an abstraction in data management that aims to overcome some of the problems encountered in data integration system. In our case it is block vector for heterogeneous data representation.
Traditionally, data integration and data exchange systems have aimed to offer many of the purported services of dataspace systems. Dataspaces can be viewed as a next step in the evolution of data integration architectures, but are distinct from current data integration systems in the following way. Data integration systems require semantic integration before any services can be provided. Hence, although there is not a single schema to which all the data conforms and the data resides in a multitude of host systems, the data integration system knows the precise relationships between the terms used in each schema. As a result, significant up-front effort is required in order to set up a data integration system.
For realization of data integration from different sources we used SQL Server Integration Services, SSIS.
For developing the portal as an architectural pattern there is used pattern Model-View-Controller (MVC).
There is specifics debug operation data space as a complex system. The query translator in Backus/Naur Form is give.

## INTRODUCTION

Information boom led to an increase in the amount of data accumulated in many subject areas thousands of times. Number of information gathered grows exponentially. Thus, according to research IDC Digital Universe Study, conducted and commissioned by EMC, the total amount of global data in 2005 was 130 Exabyte, by 2011 it increased to 1227 EB, and over the last year doubled again, reaching 3 ZB. Weather by that same survey shows that by 2015 the volume of digital data will grow to 7.9 ZB. The size of individual databases is growing very fast and overcame barrier in PB. Most of the data collected are not currently analyzed, or is only superficial analysis [15].

The main problems that arise in the data processing is the lack of analytical methods suitable for use because of their diverse nature the need for human resources to support the process of data analysis, high computational complexity of existing algorithms for analysis and rapid growth of data collected [21]. They lead to a permanent increase in analysis time, even with regular updating of hardware servers and also arise need to work with distributed database capabilities which most of the existing methods of data analysis is not used effectively. Thus, the challenge is the development of effective data analysis methods that can be applied to distributed databases in different domains. It is therefore advisable to develop methods and tools for data consolidation and use them for analysis.

Big Data information technology is the set of methods and means of processing different types of structured and unstructured dynamic large amounts of data for their analysis and use for decision support [7 - 11]. There is an alternative to traditional database management systems and solutions class Business Intelligence. This class attribute of parallel data processing (NoSQL, algorithms MapReduce, Hadoop) [1, 2, 15, 16].

Defining characteristic for big data is the amount (volume, in terms of value of the physical volume), speed (velocity in terms of both growth rate and need high-speed processing and the results), diversity (variety, in terms of the possibility of simultaneous processing of different types of structured and semi-structured data) [12 -14].

Data space is a block vector comprising a plurality of information products subject divided into three categories: structured data (databases, datawarehouses), semi-structured data (XML, spreadsheets) and unstructured data (text). It consists of the set of information products of subject area [1]. Above this vector and its individual elements defined operations and predicates that provide: converting various elements of the vector at each other; association of one type; Search items by keyword.

In thesis there is projected dataspace architecture as information technologies for working with big data.

## MATERIALS AND METHODS
## DATA SPACE ARCHITECTURE

Dataspace architecture consists of several levels and has such levels as data level, manage level and metadescribe level (Fig . 1).

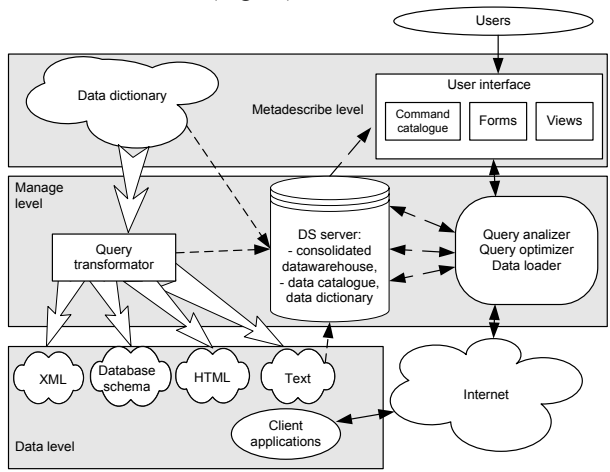

**Fig. 1**. Dataspace architecture

The modules structure of dataspace is described on Fig. 2.

Data level consists of information products of dataspace. Data level on Fig. 2 described as cloud.

Manage level consists of modules for dataspace organization and manage [1,2]:

– Module for user permission determination (by user authorized procedure).

– Query transformation module (by interpretation method).

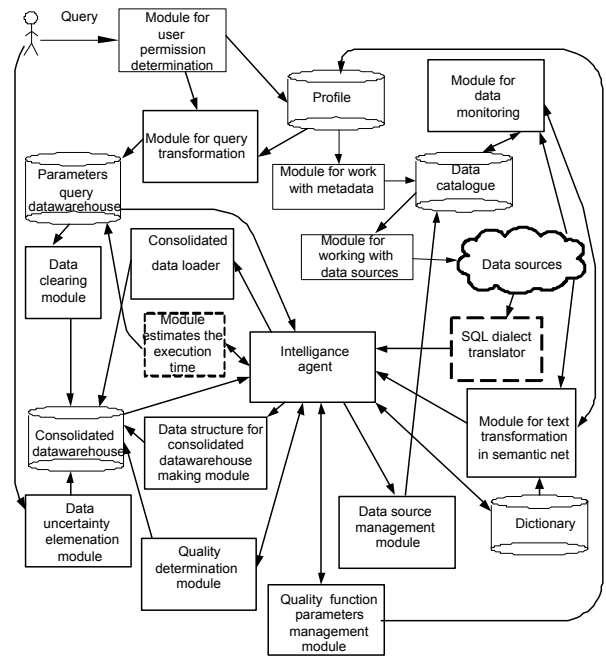

**Fig. 2**. Dataspace structure

– Module for working with metadata (by find operation as query to metadata).

– Sources access by type module (by standard data exchange protocols usage).

– Module for text transformation in semantic net (by the semi-structured data analysis method).

– Intelligent agent (based on the formal description of intelligent agent determine the structure of the data source, the algorithm of the intelligent agent).

– Data structure for consolidated datawarehouse making module (based on the method of construction of consolidated data repository schemes and work smart agent determine the structure of the data source).

– Consolidated data loader (based operations consolidation, data consolidation method).

– Module purification data (based on advanced operators cut, coagulation operator, method of forming a system of norms and criteria, method of analysis, filtering and converting input data).

– Data uncertainty elimination module (based on the method of application of classification rules and modified operator eliminate uncertainty in the network structure of the consolidated data. The method of construction schemes consolidated data repository and work smart agent determine the structure of the data source).

– Quality determination module.

– Quality function parameters management module (based methods control elements data space based on the function of the quality and levels of trust).

– Data source management module.

− Module for data monitoring.

− Module estimates the execution time (based on the standard of fixing runtime)SQL dialect translator (by the SQL description).

Level control models of platform is maintenance DS.

The meta descriptions level containing all the basic information about the data sources and methods to access them. Also there are defined methods of data processing: for structured sources - selection, grouping, etc., for semi-structured and unstructured - definition of structure or search by keyword.

In addition, the dataspace also provides data storage for storing user profiles and temporary storage request parameters.

TECHNOLOGICAL ASPECTS

For realization of data integration from different sources we used SQL Server Integration Services, SSIS. SSIS has a flexible and scalable architecture that provides effective data integration in today's business environment.

SSIS consists of the support tasks thread kernel and kernel support for the data stream. The support tasks thread kernel is oriented on operations. The flow of data exists in the context of the total flow problems (fig. 3).

The core of SSIS is the pipeline data conversion. The architecture of the pipeline supports buffering, that allowing the conveyor to quickly work with the manipulation of the data sets once they are loaded into memory. The approach is to perform all phases of ETL-process of converting data in one operation without intermediate storage. Although the specific requirements for the conversion of operations or conservation are can be an obstacle in the implementation of this approach.

SSIS if possible even avoid copying data in memory. This is fundamentally different from traditional ETL-tools that often require intermediate storage at almost every stage of processing and data integration. SSIS transforms all data types (structured, unstructured, XML, etc.) before loading into their buffers into a relational structure.

Service Integration SQL Server 2012 is optimized for connections via ADO.NET (previous versions were optimized for OLE DB or ODBC). ADO.NET using simplifies system integration and support of third parties. Integration Services SQL Server 2005 used OLE DB to perform important tasks such as search operations (lookups), but now for all tasks associated with data access, you can use ADO NET.

As the scale of integration solutions often productivity increases only to a certain limit, and then goes to a level that is very difficult to overcome. Integration Services SQL Server 2012 removes this limitation by sharing streams (threads) set of components, which increases the degree of concurrency and reduces the frequency lock, it enhances productivity in large-scale systems with a high degree of parallelization based on multiprocessor and multicore hardware platforms.

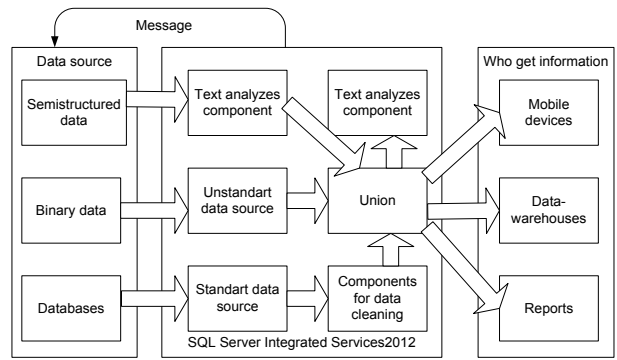

**Fig. 3**. SSIS integration schema

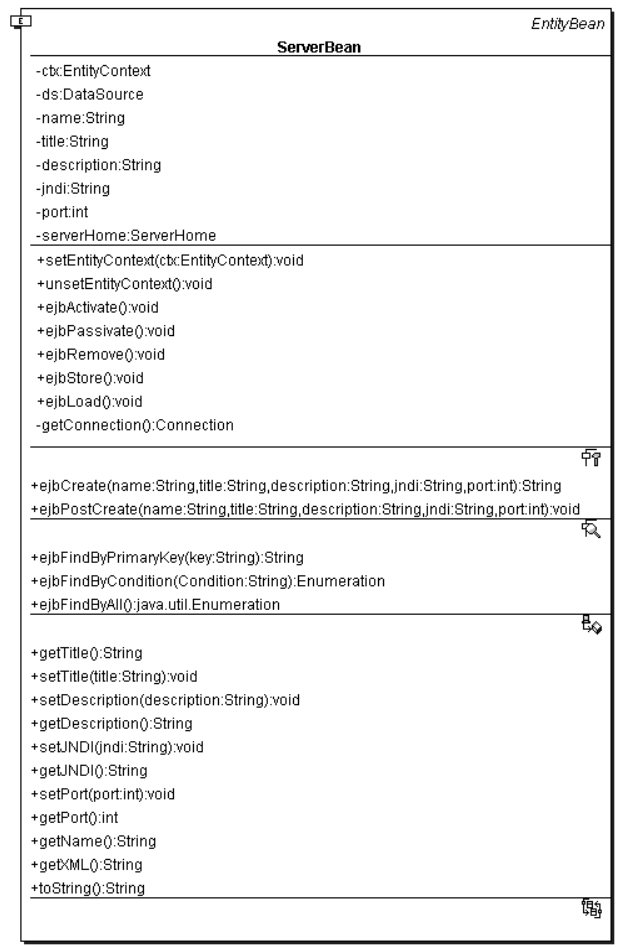

**Fig . 4**. DS components manage class structure

Search is one of the most common operations in the integration solution. Integration Services SQL Server 2008 accelerates the search operation and effectively implement them in large tables. There is loading full cache from any source, cache size can not exceed 4GB, even in 32-bit operating system. Using partial cache service integration SQL Server 2012 pre-load data required for the search. Partial cache supports OLEDB, ADO.NET and ODBC for database search, and tracks hit and miss in the search process [17–18].

SSIS can extract (and unload) data from various sources, including OLEDB, controlled sources (ADO.NET), ODBC, flat files, Excel and XML, with a special set of components called adapters. SSIS can also be used for custom adapters. It means that they are created by yourself or other manufacturers for their needs. This can include inherited logic upload data directly to the data source, which, in turn, without additional steps can be implemented in a data stream SSIS. SSIS includes a set of data conversion, with which you can do with all the data manipulations that are needed to build consolidated data repository.

THE MANAGEMENT CLASS OF DATASPACE
COMPONENT STRUCTURE

Let us describe the structure of class management components of PD (Fig. 4):

–   ctx – a reference to the object that allows a component to obtain proprietary information about users and transaction data that a user works with the component,

–   ds – reference to the pool of database connections,

–   name, title, description, jndi, port – component parameters accessible via Remote-interface methods,

–   serverHome – link to Home-interface component Server,

–   setEntityContext / unsetEntityContext - methods which establish ctx.– Invoked only container,

–   ejbActivate / ejbPassivate – methods that control life cycle component.– Invoked only container,

–   ejbRemove – a method that is called before the destruction of the component on the server side (implementing a database query to remove this component from the base),

–   getConnection – a method that cause for connection pool connections.– Its more as a convenience, and the EJB specification does not in any way,

–   ejbCreate – a method that implements a create-methods with Home-interface.– It implement database queries to create the component and set the parameters component,

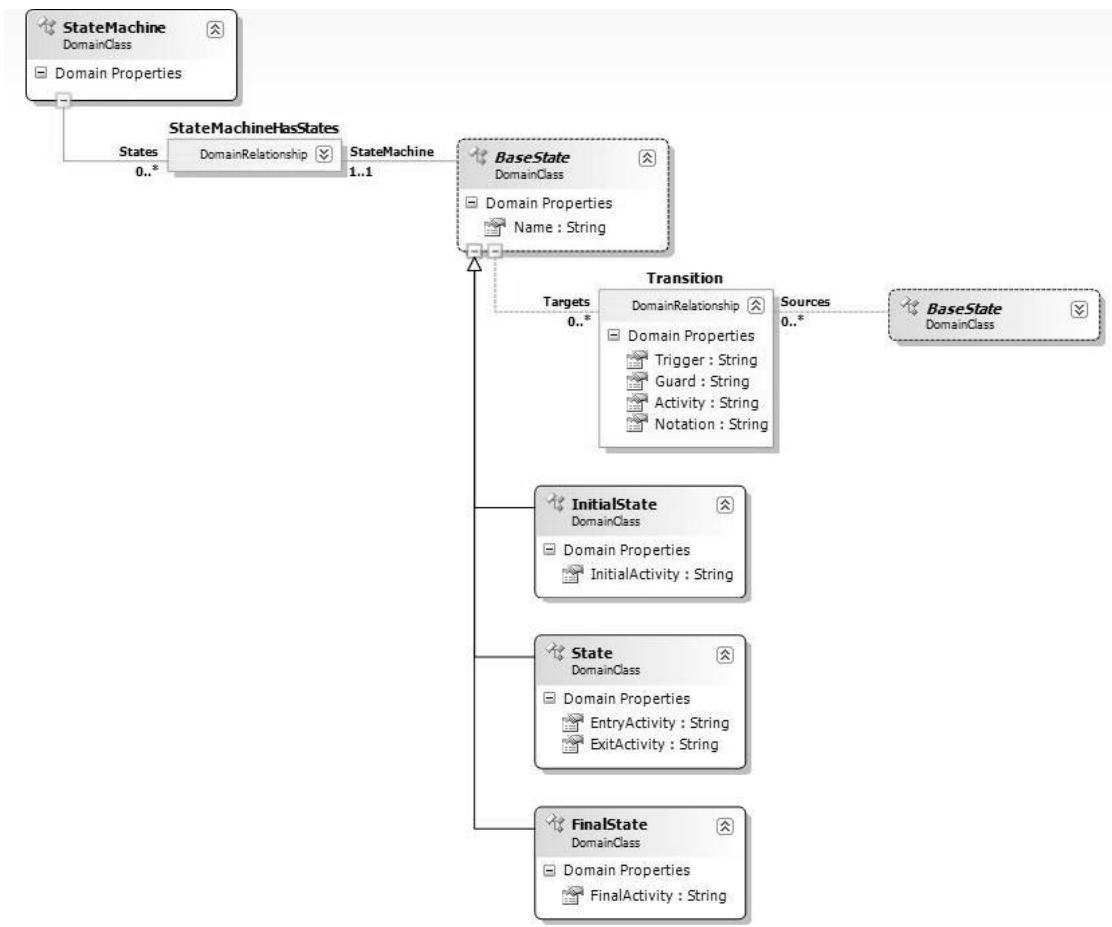

**Fig. 6**. Intelligence agent metamodel diagram

– ejbPostCreate – methods are called after ejbCreate,

– ejbFind – implement method of search techniques is searched components in the database,

– get / set – methods of implementing get / set methods defined in the Remote-interface.

– toString – defined for greater compatibility with infrastructure JAVA.

Intelligence agent metamodel is described on Fig. 5.

The root element of the metamodel is itself diagram intelligent agent [1] StateMachine. StateMachineHasStates ratio means that the agent is in the states. BaseState - base type for the state. The chart can accommodate three types of states: initial state InitialState, State intermediate state and final state FinalState. Any intermediate state can have two actions: EntryAction - action to be executed immediately at the entrance to this state, ExitAction - action to be executed upon exit from the state. The initial and final states have the single action. For the initial state of this action InitialActivity. It will be executed when you run the agent. To the end of this action FinalActivity. It will be performed at the end of intelligent agent.

Attitude Transition means a transition from one state to another.

Each transition has an event name Trigger, while the emergence of which is next. Tape Guard delivers a covenant enforcement is necessary to complete the transition. In the case of the availability and performance of necessary conditions, the proposed operation will Activity. Notation-tape, which automatically generates and submits a complete description of the transition in a format Trigger [Guard] / Action.

The main class for working with data sources is the class Model (specification is shown in Figure 7). It designed for requirements [4].

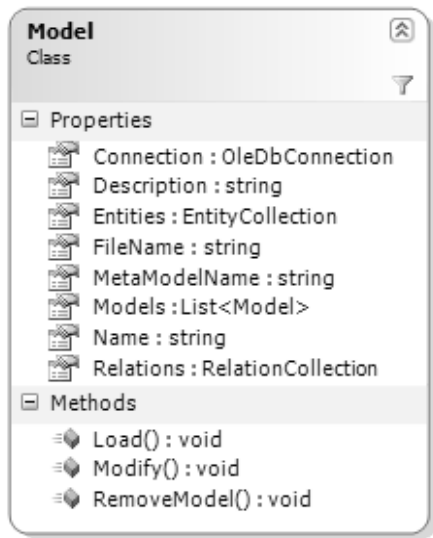

**Fig. 7**. Class Model specification

This class has such methods and properties as:

– Connection – link on data source connection.

– Description.

– FileName.

– MetaModelName – data source type.

– Models – list of linked sources.

– Name – source name.

– Entities.

– Relations.

– Load – the method for data structure load in data catalogue and data dictionary.

– Modify – the method for modify of elements in catalogue and dictionary.

– RemoveModel – the method for removing information about data source from catalogue.

EntityCollection class is presented on Fig. 8.

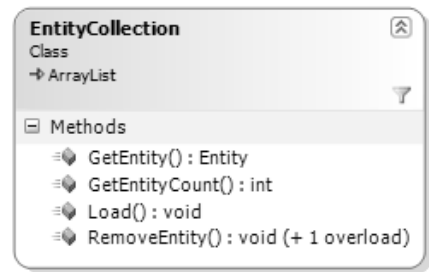

**Fig. 8**. Class EntityCollection specification

Class EntityCollection has such methods as:

– GetEntity – result of this method is element from entity collection by name.

– GetEntityCount – method for entities count in current model.

– Load – method for entity collection load from current database model.

– RemoveEntity – method for entity removing from collection.

Fig. 9. represents the class Entity.

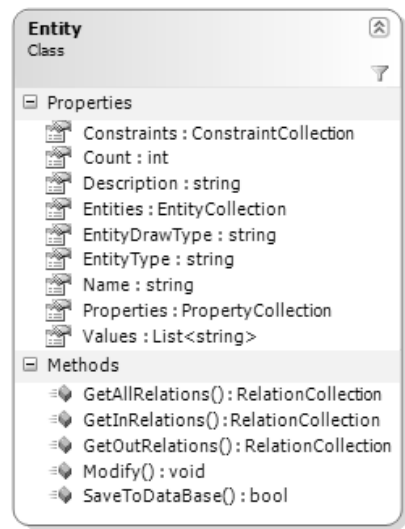

**Fig. 9**. Class Entity representation

Class Entity consists of following methods and properties:

– Attributes.

– Constraints.

– Count – count of entity elements created in model;.

– Description.

– Name.

– EntityDrawType – pictogram for entity representation.

– Entities – the model's entities collection, which consists of current entity.

– EntityType.

– Operations – list of operation about entity.

– Values – list of attribute values of entity.

– GetAllRelations – the method, result which is list of all entity relations.

– GetInRelations – the method, result which is the list of all entity get in relations.

– GetOutRelations – the method, result which is the list of all entity get out relations.

– Modify – the method for entity changing in database.

– SaveToDataBase – the method for entity saving in database.

Class Relation is given on Fig. 10. It consists of following properties:

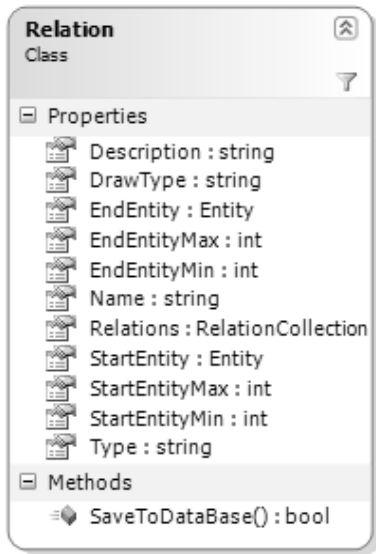

**Fig. 10**. Class Relation specification

– Constraints.

– Description.

– EndEntity – link on end entity.

– EndEntityMax – the maximum count of end entity examples.

– EndEntitytMin – the minimum count of end entity examples.

– Name.

– Relations – relations collection of current entity.

– StartEntity – link on start entity.

– StartEntityMax – the maximum count of start entity examples.

– StartEntityMin – the minimum count of start entity examples.

– Type – relation type.

The method *SaveToDataBase* saves the relation in data catalogue.

The class RelationCollection describes the entities collection (Fig.11).

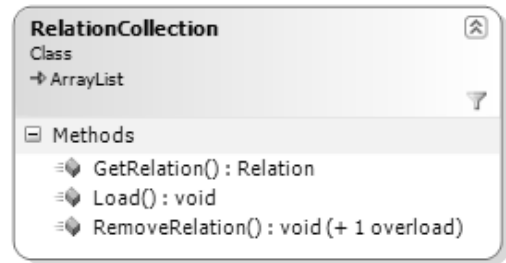

**Fig. 11.** Class RelationCollection specification

Class RelationCollection has following methods:

– GetRelation – result of this method is element of entity collection by name.

– Load – the method for entity load from current data source.

– RemoveRelation – the method for entity removing from collection and data catalogue.

The concept of "Attribute" describes a class Attribute. Specifications class is shown in Fig. 12 properties and methods of the class are:

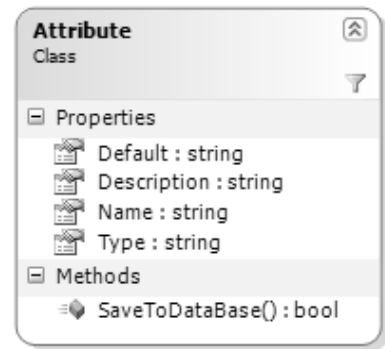

**Fig. 12**. Class Attribute specification

– Default – default value.

– Description – the attribute description.

– Name – the attribute name.

– Type – the attribute type. It may be referring to the domain of valid values, or a link to some substance.

– SaveToDataBase – the method, that is responsible for maintaining the attribute in the data directory.

A collection of attributes defined entity class AttributeCollection. Description of the class is represented in Figure 13.

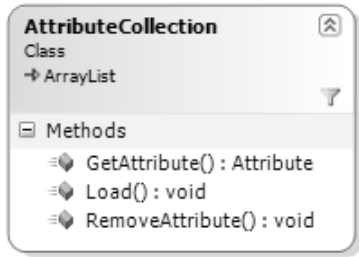

**Fig. 13.** AttributeCollection Class specification

Class AttributeCollection given by the following methods:

− GetAttribute –– a method that returns the collection of attributes to its name.

− Load –– method responsible for loading collection attributes the current nature of the source.

− RemoveAttribute –– method responsible for removing the attribute from the collection and data directory.

Class specification is shown in Fig. 14, describes the concept of "limit".

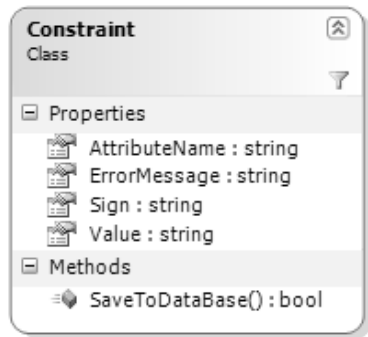

**Fig. 14.** Class Constraint specification

This class has following properties:

− ErrorMessage – error message on constraint.

− AttributeName – name of attribute.

− Sign – constraint signature.

− Value – value in right part of constraint.

The *SaveToDataBase* method saves information about constraint in database (data catalogue).

Constraint collection for entities and relations is presented by ConstraintCollection class. This class description is presented on Fig. 15.

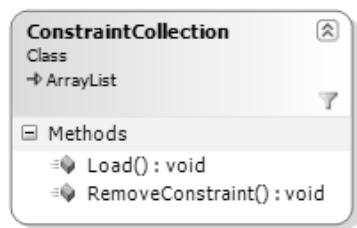

**Fig. 15.** Class ConstraintCollection specification

The methods of ConstraintCollection class are:

− Load – the method for constraint loading from collection.

− RemoveConstraint – the method for constraint removing from collection and dataspace.

LANGUAGE ELEMENTS DESCRIPTION

For translator building we must describe elements of query language to dataspace. We used Backus/Naur Form, BNF [19 − 20].

<letter>::=a|b|c|d|e|f|g|h|i|j|k|l|n|m|o|p|q|r|s|t|u|v|w|x|y|z|A|B|C|D| E|F|G|H|I|J|K|L|N|M|O|P|Q|R|S|T|U|V|W|X|Y|Z

<keyword> ::= (<keyword>) |<letter> | < keyword>

<number> ::= 0|1|2|3|4|5|6|7|8|9

<object> ::= <data catalogue element>

<par> ::= <the synonym of data catalogue element >

<param> ::= <keyword>[{<keyword>|<number>}]

<num> ::= <number>[{<number>}]

<expr> ::= <operand> [{<op> <operand>}]

<operand> ::=» («<expr>»)» | <num> | <param> [«[«<expr>»]»]

<op> ::= <grteq>

<inv> ::= <logicalop> | «*» | «/»

<type> ::= «SUM» | «COUNT» | «AVG»

<logicalop>::= «<» | «>» | «>=» | «<=» | «=» | «<>» | [<op>]

<whereop> ::= «where» «(» <object> [«:» <par>] {«,»<object> [«:» <par>] }«)»

<whoop> ::= «who» «(» <object> [«:» <par>] {«,»<object> [«:» <par>] } «)»

<howop> ::= «how» «(» <object> [«:» <par>] {«,»<object> [«:» <par>] } «)»

<Seop> ::= «Se» «(»<object>[«:»<par>] [«Agg»<type>] {«,» <object> [«:» <par>] [«Agg» <type>] «)»

<whatop> ::= «what» «(» <object> [«:» <par>] {«,»<object> [«:» <par>] ]«)»

<whichop>::= «which» «(» <object> [«:» <par>] {«,»<object> [«:» <par>] ]«)»

<Semantop>::= «Semant» «(» <object> [«,» <object> ]«)»

<Consop> ::= «Cons» «(»<object> [«:» { <par> <operator> <param>}] «)»

<profileop> ::= «where» «(» <object> [«:» <num>] {«,»<object> [«:» <num>] }«)»

<Unionop> ::= «Union» «(» <object> [«:» <par>] {«,»<object> [«:» <par>] } «)»

<Uniolop> ::= «Union» «(» <object> [«:» <par>] {«,»<object> [«:» <par>] } «)»

<Intersop> ::= «Inters» «(» <object> [«:» <par>] {«,»<object> [«:» <par>] } «)»

< Differop> ::= «Differ» «(» <object> [«:» <par>] {«,»<object> [«:» <par>] } «)»

INTERFACE REALIZATION

Let us project interface metamodel for user query interpretation (Fig. 16). Entities InterfaceHasMethods, InterfaceHasProperties, InterfaceHasEvents meen, that interface has Methods, Properties and Events.

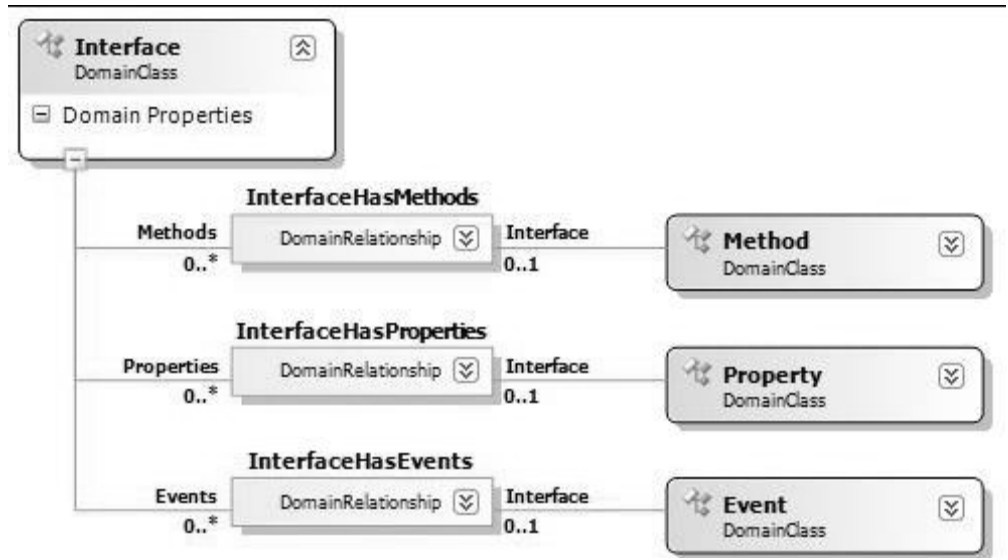

**Fig. 16**. Interface metamodel for user query interpretation

For developing the portal as an architectural pattern there is used pattern Model-View-Controller (MVC).

Model-View-Controller (Model-View-Controller, MVC) is architectural pattern, which is used in the design and development of software. Splits system into three parts: data model and data view. It is used to separate data (model) from the user interface (view) so that the user interface changes minimally affect the operation of the data, and changes in the data model could be conducted without changing the user interface.

The purpose of the template is flexible design software, which should facilitate further changes or expansion programs, and provide an opportunity for reuse of individual components of the program. Also use this template in large systems leads them in a certain order and makes clearer by reducing their complexity.

The architectural pattern Model-View-Controller (MVC) divides the program into three parts. In the triad of responsibilities Component Model (Model) is a data storage and software interface to them. View (View) is responsible for the presentation of these data to the user. Controller (Controller) manages components, receiving signals as a response to user actions, and reporting changes-component model.

Model encapsulates core data and basic functionality of their treatment. Also component model does not depend on the process input or output. Component output view can have several interconnected domains, such as various tables and form fields, in which information is displayed. The functions of the controller is monitoring the developments resulting from user actions (change of the mouse, pressing buttons or entering data in a text field).

Registered events are shown in different requests that are sent to the component models or objects responsible for displaying data. Separation of models from data presentation allows independent use different components to display. Thus, if the user through controller makes a change in the data model, the information provided by one or more visual components will be automatically corrected according to the changes that have occurred.

At the level Model used ORM (Object-relational mapping), including technology Entity Framework. At this level creates a database model that allows you to work with it as with a set of entities, as well as avoiding explicit use of SQL. All these things will perform ORM.

Controller is a class that contains event handlers and other business logic.

## CONCLUSIONS

1. In this paper there is projected dataspace architecture and instrumentation tools for practical realization.

2. There are chased program tools for variant data integration realization.

3. The main classes' specification is described.

4. There are described language tools and user interface realization.